\documentclass[amssymb,prb,twocolumn,showpacs,superscriptaddress]{revtex4}
\usepackage{amsmath}
\usepackage{graphicx}
\usepackage{verbatim}
\usepackage{graphics}
\usepackage{units}
\usepackage{color}
\usepackage{ulem}
\providecommand{\tabularnewline}{\\}

\makeatother

\usepackage[english]{babel}
\begin{document}

\title{Quasiparticles in Iron Silicides:\\ GW {\it versus} LDA}

%\author{{ Natalie Zamkova$^{1,2}$, Vyacheslav Zhandun$^{1}$,\\ Sergey Ovchinnikov$^{1,2}$,
%and Igor Sandalov$^{1,3}$ \\
%{\it $^{1}$L.V.Kirensky Institute of physics,\\ Siberian Branch of Russian
%	Academy Sciences,\\ 660036 Krasnoyarsk, Russia;\\
%	 $^{2}$ Siberian Federal University, 660041 Krasnoyarsk, Russia;\\
%	 $^{3}$ Applied Materials Physics, Department of Materials Science and Engineering, KTH Royal Institute of Technology, SE 100 44 Stockholm, Sweden}}}

%\maketitle
\author{Natalia Zamkova}
\affiliation{L.V.Kirensky Institute of physics, Siberian Branch of Russian Academy Sciences, 660036 Krasnoyarsk, Russia}
\affiliation{Siberian Federal University, 660041 Krasnoyarsk, Russia}
\author{Vyacheslav Zhandun}
\affiliation{L.V.Kirensky Institute of physics, Siberian Branch of Russian Academy Sciences, 660036 Krasnoyarsk, Russia}
\author{Sergey Ovchinnikov}
\affiliation{L.V.Kirensky Institute of physics, Siberian Branch of Russian Academy Sciences, 660036 Krasnoyarsk, Russia}
\affiliation{Siberian Federal University, 660041 Krasnoyarsk, Russia}
\author{Igor Sandalov}
\email{sandalov@kth.se}
\thanks{Corresponding author}
\affiliation{L.V.Kirensky Institute of physics, Siberian Branch of Russian Academy Sciences, 660036 Krasnoyarsk, Russia}
\affiliation{Applied Materials Physics, Department of Materials Science and Engineering, KTH Royal Institute of Technology, SE 100 44 Stockholm, Sweden}

\begin{abstract}
The angle-resolved photoemission spectroscopy (ARPES) is able to measure
both the spectra and spectral weights of the quasiparticles in solids.
Although it is common to interpret the band structure obtained within
the density-functional-theory based methods as quasiparticle spectra,
these methods are not able to provide the changes in spectral weights
of the electron excitations. We use Vienna Ab initio Simulation Package
(VASP) for evaluation of the quasiparticle spectra and their spectral
weights within Hedin's GW approximation (GWA) for $Fe_{3}Si$ and
$\alpha-FeSi_{2}$, providing, thus, a prediction for the ARPES experiments.
Comparison of the GGA-to-DFT and GWA band structures shows that both
theories reflect peculiarities of the crystal structures in similar
shape of the bands in certain $k$-directions, however, in general
the difference is quite noticeable. We find that the GWA spectral
weight of quasiparticles in these compounds deviates from unity everywhere
and shows non-monotonic behavior in those parts of bands where the
delocalized states contribute to their formation. Both methods
lead to the same conclusion:  those of iron ions in $Fe_{3}Si$ which occupy the positions $Fe^{(1)}$, where
they are surrounded by only $Fe$ ions ( $Fe^{(2)}$ positions), have the  $d$-electrons localized and
large magnetic moment whereas $Fe$ ions in the $Fe^{(2)}$ positions with
the $Si$ nearest neighbors have $d$-electrons delocalized and the magnetic
moments quenched. The $Si$ influence on the $Fe$ ion state is even more pronounced in 
the $\alpha-FeSi_{2}$ where all iron ions have the $Si$ ions as nearest neighbors: 
 both GWA and GGA calculations produce zero moment on iron ions. 
 
The advantages and disadvantages of both approaches are discussed.
\end{abstract}

\maketitle

\section{Introduction}

A hope to use the electron spin additionally to the charge as an information
carrier has led to a development of the spintronics. One feasible
way to exploit the spin degrees of freedom is to synthesize such magnetic
semiconductors, which, on the one hand, should be magnetic at room
temperature, and, on the another hand, should be easily integrated
with existing semiconductor industry. Therefore, it is desirable that
it should be $Si$ based \cite{key-1}. A magnetic moment can be added by
a transition-metal constituents. The other way is to use a magnetic
metal for injecting of spin-polarized electrons into, say, $Si$-based
semiconductor. The technologies which create magnetic epitaxial multi-layer
films on $Si$, produce an interface, which contains the compounds
of $TSi$, where $T$ is a transition metal. This makes iron silicides
compounds highly perspective materials both in bulk and film form
and a detailed understanding of their physics is on demand. Recently
the formation of single crystalline$Fe_{3}Si$ phases in the $Fe/Si$
interface has been demonstrated by several groups \cite{key-2}-\cite{key-4}.
The theoretical understanding of the ground-state properties like
cohesive and structural properties is achieved long ago via first-principle
calculations based on the various realization of local-density approximation
to density functional theory \cite{key-5}-\cite{key-10}.

However, the experiments, that measure differential information (not
thermodynamics), require for their interpretation a knowledge of the
single-particle-{\it excitation} properties. For example, all photoeffect-based
measurements belong to this class. The photoemission spectroscopy
(\textbf{PES}) provides direct measurement of the energy of electronic
quasiparticles. Its extension, the angle-resolved PES (\textbf{ARPES}),
which uses the synchrotron radiation facilities, allows for extracting
also perpendicular-to surface momentum dependence of quasiparticle
energy. Further refinement, the laser-based ARPES provides even better
accuracy and resolution. Recently developed method, the time-resolved
two-photon photoemission (\textbf{TR-2PPE}) spectroscopy \cite{key-11},
can monitor the state of an excited electron during the course of
its transformation by laser-induced surface reaction. Probably, this
is the only method capable to directly measure the quasi-particle's
life-time. The transport and tunneling experiments are even more evident
examples where the qusiparticle concept is a necessary ingredient
for understanding the underlying physics. However, general theories
sometimes are not sufficient for describing real materials. For example,
the predictions of the life-times within the Landau theory of Fermi-liquid
do not describe the experiments even on $Al$ and noble metals\cite{key-12},
contrary to the expectation. Indeed, electrons in these metals are
well delocalized and expected to behave as a good Fermi-liquid. Furthermore,
TR-2PPE experiments show that the lifetime of an excited electron
in $Al$ at a fixed energy $E<E_{F}$ depends on the frequency of
the pump pulse, i.e., on the band from which the electron originated.
These examples convince that {\it ab initio} calculations for real
materials are required. The most developed method of electronic structure
calculations, which is based on density functional theory, is designed
and applicable for the ground state properties only {\cite{key-12}},{\cite{key-18b}}.
This gives rise to certain doubts that the electronic bands obtained
within the framework density functional theory (DFT) by means of local-(spin)-density
approximation with or without gradient corrections, can be interpreted
as energies of excited states. Nevertheless, the band structure, generated
by Kohn-Sham equations, is ubiquitously  exploited as quasiparticle
spectrum.Here we will compare the results of calculations
within the version of gradient-corrected local-density approximation
to density functional theory (GGA) and GW approximation (GWA) \cite{key-12}.
In the latter $G$ is the Green's function and $W$ is the screened
Coulomb interaction. Both approaches have their own advantages and
disadvantages. The exchange-correlation potential used in all modern
implementaions of the Kohn-Sham machinery contains the correlation
effects that are much beyond the random phase approximation used in
GWA. Nevertheless the DFT based approach has the essential drawbacks
that i) the exchange-correlation potential is calculated for the homogeneous
electron gas, and the transfer of these results to the inhomogeneous
electron gas of real materals is not easy to justify, and ii) it is
difficult to impove the calculations by adding in a controlled way
some corrections. The convincing example of it is the widely used
LDA+U approximation, where many different forms of double-counting
corrections are in use: the way to make it in a controlled way is
not found yet. The GW method does not contain this problem, it is well controlled
approximation. The strong advantage of GWA is that it results in real
quasiparticle excitations. However, it requires much more of computer
resources than, say, GGA. It is expected that they should produce
approximately the same band structure, since both approaches ignore
the intraatomic (often strong) correlations. If the problem of interest
does not require the knowledge of the spectral weights and/or is
intended to use it as a first step for a better approximations, it
is preferable to use GGA. The latter usually is used as a perturbational
input for GWA. For this reason it makes sense to compare both quasiparticle
and Kohn-Sham band structures and to analyse the difference between
the exchange-correlation potential and the real part of the GW self
energy. Since the eigenvalues of the Kohn-Sham (KS) equations cannot
be interpreted as quasi-particle (QP) excitations whereas the band
structure generated by the GW method can, we calculate GW QP energies.
They can serve as a predicton of ARPES measurements on iron silicides. 

The paper is organized as follows. In Sec. II we present the details
of our {\it ab initio} DFT and GW calculations and give brief discussion
of their advantages and drawbacks. The comparison of band structure
and densitiy of electron states calculations of iron silicides $Fe_{3}Si$
and $\alpha-FeSi_{2}$ obtained by both methods is given in Sec. III
and in Sec. IV we discuss results and make conclusions.

\section{The Methods: GW versus GGA for iron silicides}

\subsection{Band structure and excitation energies\label{sub:BandStr_vs_Excit}. }

The question if the DFT band structure can be interpreted as single-electron
excitations' energies have been discussed repeatedly (see, {\it e.g.},\cite{key-5}, review \cite{key-18a} 
and the book \cite{key-18b}).
The single-electron Green's function by definition describes the transitions
between $n$ and $\left(n\pm1\right)$ electron states, and the solutions
of the Dyson's equation provide the excitation energies and their
spectral weights. The potential in the Kohn-Sham equation does not
depend on energy and automatically looses information on possible
deviations of the spectral weight of quasi-particles from unity. Although
 the DFT based methods have been
designed for the investigation of the ground-state properties and,
it seems, there are no grounds for interpreting the band structure,
generated by the Kohn-Sham equations as the energies of the single-electron
excitations. However, the question about such a possibility is not
so trivial and cannot be ruled out by this argument only. First, one
can notice that the same statement holds for the Hartree-Fock-determinant
wave function of the {\it ground state}, for which Koopmans 
\cite{key-19} proved that in closed-shell Hartree - Fock
theory, the first ionization energy of a molecular system is equal
to the negative of the orbital energy of the highest occupied molecular
orbital. There is an analogue of the Koopmans' theorem for DFT \cite{key-20}.
Second, Hoenberg-and-Kohn's proof \cite{key-21}  that the total
energy of a system with interaction is a unique functional of the
electron density $\rho({\bf r})$ led Sham and Schlueter \cite{key-22}
to the equation, which, at least, in principle, allows calculation
of the exchange-correlation potential in DFT for any chosen approximation
for the electron-Green's-function self-energy $\Sigma'$. Schematically, 
$ G^{-1}  =  (G_{0}^{-1}-v_{xc})+v_{xc}-\Sigma' $ gives 
$ G  =  G_{DFT}+G_{DFT}\left(v_{xc}-\Sigma'\right)G.$ 
Here $G_{0}$ is the Green's function of electrons in the Coulomb field
of nuclei and Hartree potential and the prime in $\Sigma'$ means
that the Hartree term is subtracted, $\Sigma'=\Sigma-v_{Hartree}$.
Expressing then the charge densities in terms of $G$,  $\rho=\left\langle G\right\rangle _{E}$,
i.e., taking appropriate contour integral over energy from both sides
and using the equality $\rho_{model}^{DFT}(r)=\rho_{genuine}(r)$
one finds $0=\left\langle G_{DFT}\left(v_{xc}-\Sigma'\right)G\right\rangle _{E}.$
This equation \cite{key-22} establishes the correspondence between the
self-energy and the exchange-correlation potential. For this reason one can hope that at least
for the $ v_{xc}$ found for certain $\Sigma' $ the 
self-energy-based and Kohn-Sham eigenvalues coincide.  The problem is, however,
 that the eigenvalues $\varepsilon_{kn}$ and wave functions $\varphi_{kn}(r)$ of
Kohn-Sham equation,
\begin{widetext}
\begin{equation}
\left[\frac{p^{2}}{2m}+v_{e-n}(r)+v_{Hartree}(r)+v_{xc}(r)\right]\varphi_{kn}(r)=\varepsilon_{kn}\varphi_{kn}(r),\label{eq: Kohn-Sham-1}
\end{equation}
and its self-energy-based counterpart,
\begin{equation}
\left[\frac{p^{2}}{2m}+v_{e-n}(r)+v_{Hartree}(r)\right]\Phi_{kn}(r,E_{kn})+\int dr_{1}Re\Sigma'(r,r_{1},E_{kn})\Phi_{kn}(r_{1},E_{kn})=E_{kn}\Phi_{kn}(r,E_{kn}),\label{eq: eq for QP-1}
\end{equation}
\end{widetext}
 represent more differential information than the electron densities
and, to the best of our knowledge, a prove of a statement that
$\varepsilon_{kn}=E_{kn}$ does not exists ( here $n$ labels the
bands). To be more precise, Eq.(\ref{eq: eq for QP-1}) is an approximate version  
of the equation for the functions $\Phi_{kn}(r,\omega)$,
which  diagonalize the Green's function in GWA at arbitrary energy $\omega$, 
$ \left\langle \Phi_{kn}(\omega)\right|G^{-1}(\omega)\left|\Phi_{k'n'}(\omega)\right\rangle =\delta_{kk'}\delta_{nn'}\left(\omega-\Delta_{kn}\left(\omega\right)\right) $,
{\it i.e.}, it contains energy-dependent self-energy $\Sigma'(r,r_{1},\omega)$ and $\Phi_{kn}(r_{1},\omega)$; 
the eigenvalues, which are poles of the Green's function, are obtained 
from the equation $E_{kn}=Re\Delta_{kn}\left(E_{kn}\right)$.
 At last, both equations are highly non-linear and uniqueness theorem 
does not exist for them. Particularly, the problem of choice of the physical solution from 
multiple solutions generated by GWA is discussed in ref.\cite{key-31}. 
Thus, at the moment we are at the stage when even the form of the exchange-correlation potential 
for a self-energy in GW approximation is not derived and the question about relationship between 
Kohn-Sham and GWA energies remains open.
It is interesting, however, that more accurate exchange-correlation functionals
than the widely-used ones can be obtained for evaluation of the excitation energies: 
the authors of ref. \cite{key-5} state that, on the one hand, a surprising degree of agreement 
between the exact ground-state Kohn-Sham eigenvalue differences and excitation energies, for excitations
from the highest occupied orbital to the unoccupied orbitals has been
achieved for small systems; on the other hand, that ``{\it the popular
LDA and GGA functionals are nowhere near sufficiently accurate}''\cite{key-5}.

\subsection{Band structure and spectral weights\label{sub:BandStr_vs_SpectrW}.}

The PES-based experiments provide also the information on the spectral-weights
of QPs, $Z_{\varepsilon}$. As was discussed above, the ``self-energy'' in Kohn-Sham equation
(\ref{eq: Kohn-Sham-1}) does not depend on energy and, therefore,
corresponding spectral weight, defined in the perturbation theory
as $\left(Z_{k\lambda}=\left[1-\partial\Sigma(E)/\partial E\right]^{-1}\right)_{E=E_{k\lambda}}$,
is always equal to unity. Insufficiency of the DFT-based approach 
is seen also from the Sl\"{u}ter \& Sham's equation for the exchange-correlation potential: 
since it is based on the equalities of model and genuine electron densities of many-electron system, which are 
integral characteristics,  it does not provide  the equation for spectral
weights of the excitations. Contrary to that the self-energy does depend on energy 
giving birth to the spectral weight of the GW-QPs $Z_{\varepsilon}<1$;
 the remaining part of the spectral weight
is shifted to incoherent excitations. It is instructive to represent
the spectral weight in terms of magnitudes, encoded
in the VASP. Let us write the equations for the Green's functions,
corresponding to Eqs.(\ref{eq: Kohn-Sham-1}),(\ref{eq: eq for QP-1}):
\begin{equation}
G^{-1}(E)=G_{DFT}^{-1}(E)-\left(\Sigma'(E)-v_{xc}\right).\label{eq:PT over vxc-sigma}
\end{equation}
Within the ``one-shot'' ($G_{0}W_{0}$) approximation $E=\varepsilon_{kn}$
and we obtain 
\begin{equation}
Z_{kn}\simeq\frac{E_{kn}-\varepsilon_{kn}}{Re\Sigma_{GW}(\varepsilon_{kn})-\left(v_{xc}^{(GGA)}\right)_{kn}}.\label{eq:Z_kn}
\end{equation}
This relation is fulfilled  by the magnitudes $E_{kn},\varepsilon_{kn},\left(v_{xc}^{(GGA)}\right)_{kn},\Sigma_{GW}(\varepsilon_{kn}),$
which are generated by VASP. As seen from Eq.(\ref{eq:Z_kn}) in the form \cite{key-18a},
\begin{equation}
E_{kn}=\varepsilon_{kn}+Z_{kn}\left[Re\Sigma'_{GW}(\varepsilon_{kn})-\left(v_{xc}^{(GGA)}\right)_{kn}\right],\label{eq:E=00003Dpdf+Z(Sigma-Vxc)}
\end{equation}
 the role of the coefficient in the first term of the quasiparticle-energy
expansion with respect to presumably small perturbation $Re\Sigma\left(\varepsilon_{kn}\right)-v_{xc},$
is played by the spectral weight. If the perturbation theory works, 
 $E_{kn}-\varepsilon_{kn}<\left(v_{xc}-\Sigma\right)_{kn}$
everywhere.  Our calculations confirm that this inequality is valid 
for iron silicides.  A comparison of behavior
of these two differences, $k$-dependence of the spectral weights and an analysis of contributions 
from different electrons into these $k$-dependences are given in Sec.(\ref{sub:Compar-of-GGA_GW}).

\section{Iron silicides }
The calculations presented in this paper are performed using the Vienna
Ab initio Simulation Package (VASP\textbf{)} \cite{key-13} with Projector
Augmented Wave (PAW) pseudopotentials \cite{key-14}. The valence
electron configurations $3d^{6}4s^{2}$ are taken for $Fe$ atoms
and $3s^{2}3p^{2}$ for $Si$ atoms. One part of calculations is based
on the density functional theory where the exchange-correlation functional
is chosen within the Perdew-Burke-Ernzerhoff (PBE) parametrization
\cite{key-15} and the generalized gradient approximation (GGA). 

In the GW part of calculations implemented in VASP \cite{key-16}
we report only one-shot approximation, so-called $G_{0}W_{0}$. The
GGA Kohn-Sham band structure and eigenfunctions were taken as the
input for the GW calculations. The self-energy $\Sigma$ is
computed as $\Sigma\approx iG^{GGA}W^{GGA}$. Throughout
 both GGA and GWA calculations the plane wave cutoff energy is
500 eV, the Brillouin-zone integrations are performed on the grid
Monkhorst-Pack \cite{key-17} special points $10\times 10 \times 10$
for $Fe_{3}Si$ and $12\times 12\times 6$ for $\alpha-FeSi_{2}$
and Gauss broadering with smearing 0.05eV is used. For all GW calculations
the number of frequencies is 500 and 160 electronic bands are used.
We use the complex shift of 0.043 eV for the calculation of self-energy.

\subsection{Iron Silicides structure}

The calculations are performed for two iron silicides, $Fe_{3}Si$
and $\alpha-FeSi_{2}$. The compound $Fe_{3}Si$ belongs to $DO_{3}$
structural type with the space symmetry group $Fm\bar{3}m$. The iron
atoms have two nonequivalent crystallographic positions in {\it fcc}
lattice, namely, $Fe^{\left(1\right)}$and $Fe^{\left(2\right)}$ have different\textbf{
}nearest surroundings\textbf{:} $Fe^{\left(1\right)}$ has eight $Fe^{\left(2\right)}$
nearest neighbors\textbf{ }which form a cube, whereas the $Fe^{\left(2\right)}$
is in the tetrahedral surrounding of both $Si$  and $Fe^{\left(1\right)}$
atoms. 

The iron bi-silicides have several structural modifications. The most
stable phases are $\alpha-FeSi_{2}$ and $\beta-FeSi_{2}$ phases
\cite{key-22},\cite{key-23}. The compound $\alpha-FeSi_{2}$ has tetragonal lattice
with $P4/mmn$ space symmetry group with one molecula per unit cell.
Each iron atom here is located in the center of cube, consisting of
the silicon atoms. This structure contains the planes which are formed
by only iron and by only silicon atoms. These planes are orthogonal
to tetragonal axis. Two planes formed by silicon atoms are separated
by wide empty cavity, which does not contain the iron atoms. According
to our calculations this cavity is expected to play essential role
in the transport properties of $\alpha-FeSi_{2}.$ 

The rhombohedral cell have been used for the $Fe_{3}Si$ calculations.
The equilibrium parameters and the distances between nearest $Fe$
and $Si$ atoms for the $Fe_{3}Si$ and $\alpha-FeSi_{2}$ structures
have been found from the full optimization of the structure geometries
within GGA and are shown in Tab. (\ref{tab:Lattice-parameters}).
The GW calculation have been performed with the same structural parameters.

\begin{table}[htb]
\hspace{0.3cm}
\begin{tabular}{|c||c|}
\hline 
$Fe_{3}Si$  & $\alpha-FeSi_{2}$\tabularnewline
\hline 
$a=5.6\textrm{\AA}\left(5.65\textrm{\AA}\right)$ & $a=2.70\textrm{\AA}\left(2.69\textrm{\AA}\right)$\tabularnewline
$R\left(Fe^{(1)}-Fe^{(2)}\right)=2.45\textrm{\AA}$ & $c=5.13\textrm{\AA}\left(5.134\textrm{\AA}\right)$\tabularnewline
$R\left(Fe^{(2)}-Si\right)=2.45\textrm{\AA}$ & $z_{Si}=0.272\left(0.28\right)$\tabularnewline
$R\left(Fe^{(1)}-Si\right)=2.83\textrm{\AA}$ & $R\left(Fe-Si\right)=2.30\textrm{\AA}$\tabularnewline
 & $R\left(Si-Si\right)=2.56\textrm{\AA}$\tabularnewline
\hline 
\end{tabular}

\protect\caption{\label{tab:Lattice-parameters}Relaxed lattice parameters and the
equilibrium distances between nearest ions. The experimental values
\cite{key-22} are given in brackets. }
\end{table}

Both spin-polarized GGA and GW result in metallic states,  ferromagnetic  for
$Fe_{3}Si$ and  paramagnetic with zero-spin $Fe$ atoms for $\alpha-FeSi_{2}$. 
 For this reason all further calculations
for $\alpha-FeSi_{2}$ have been performed within non-spin-polarized
version of VASP. The structural inequivalence of the $Fe$-atoms'
surroundings in the $Fe_{3}Si$ reflects itself in both magnetic moment
values and the  contributions of $Fe$-ions' $d$-states into electron density
of states. The magnetic moment $M_{Fe1}$ of $Fe^{\left(1\right)}$ atom is higher
than the\textbf{ }free-atom moment, $M_{Fe1}^{GGA}=2.52\mu_{B}$ and
$M_{Fe1}^{GWA}=2.55\mu_{B}$ what corresponds approximately to the
$d^{5}$ configuration. The $Fe^{\left(2\right)}$atom has much lower
moment, $M_{Fe2}^{GGA}=1.34\mu_{B}$ and $M_{Fe2}^{GWA}=1.40\mu_{B}.$
As will be seen from the analysis of DOS, the latter moments are formed
by the delocalized $d$-states. The experimental values reported in works \cite{key-24} and \cite{key-25}  
are slightly different:
 $M_{Fe2}^{exp}=1.2\mu_{B} $, $M_{Fe1}^{exp}=2.4\mu_{B}$ in Ref.
\cite{key-24} and $M_{Fe2}^{exp}=1.35\mu_{B}$,
$M_{Fe1}^{exp}=2.2\mu_{B}$ in Ref.\cite{key-25}.

\subsection{Comparison of GGA and GW densities of electron states }

\begin{widetext}

\begin{figure}[htb]
%\hspace{3mm} 
\includegraphics[width=0.45\linewidth,clip]{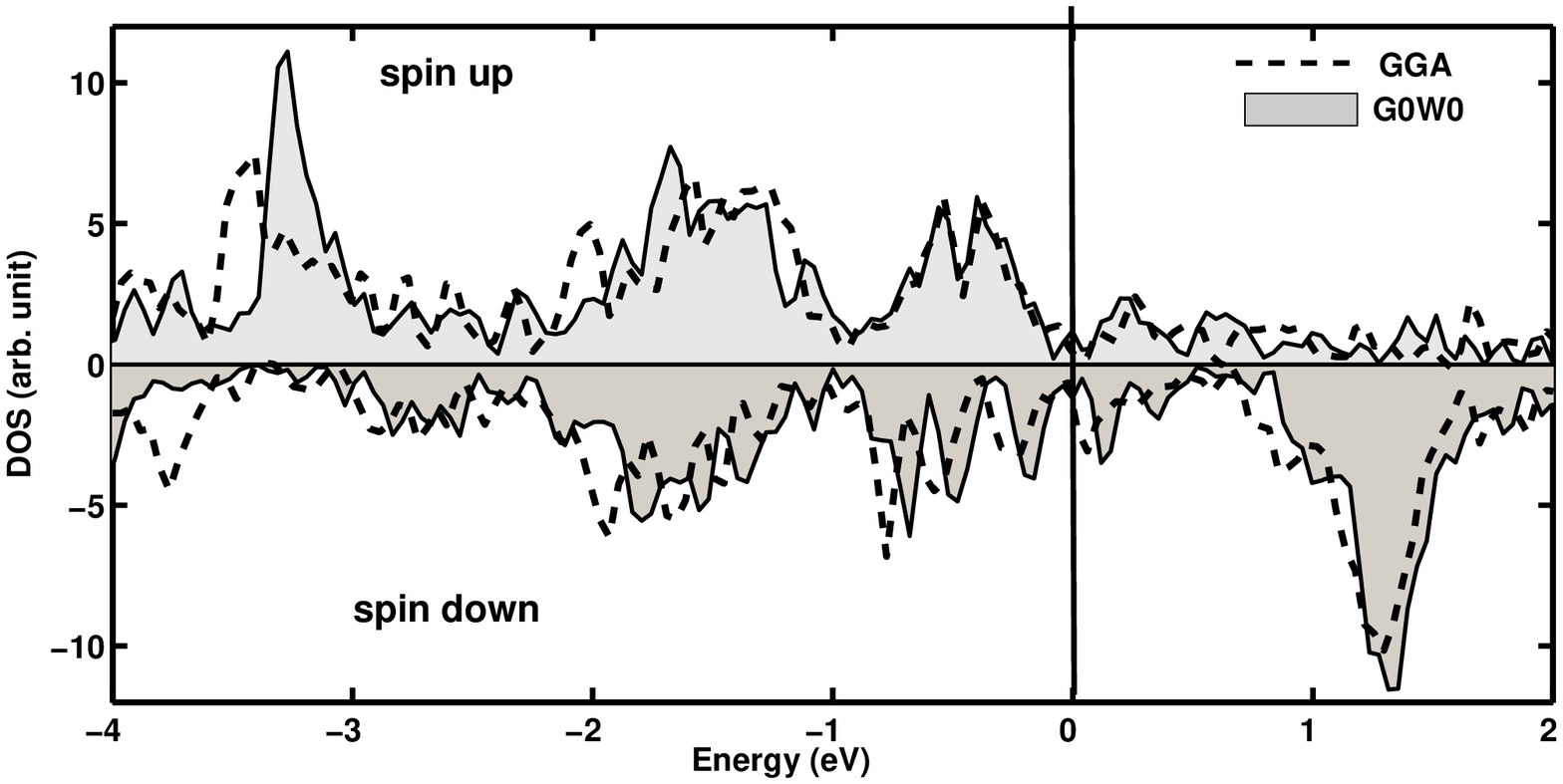} \includegraphics[width=0.45\linewidth,clip]{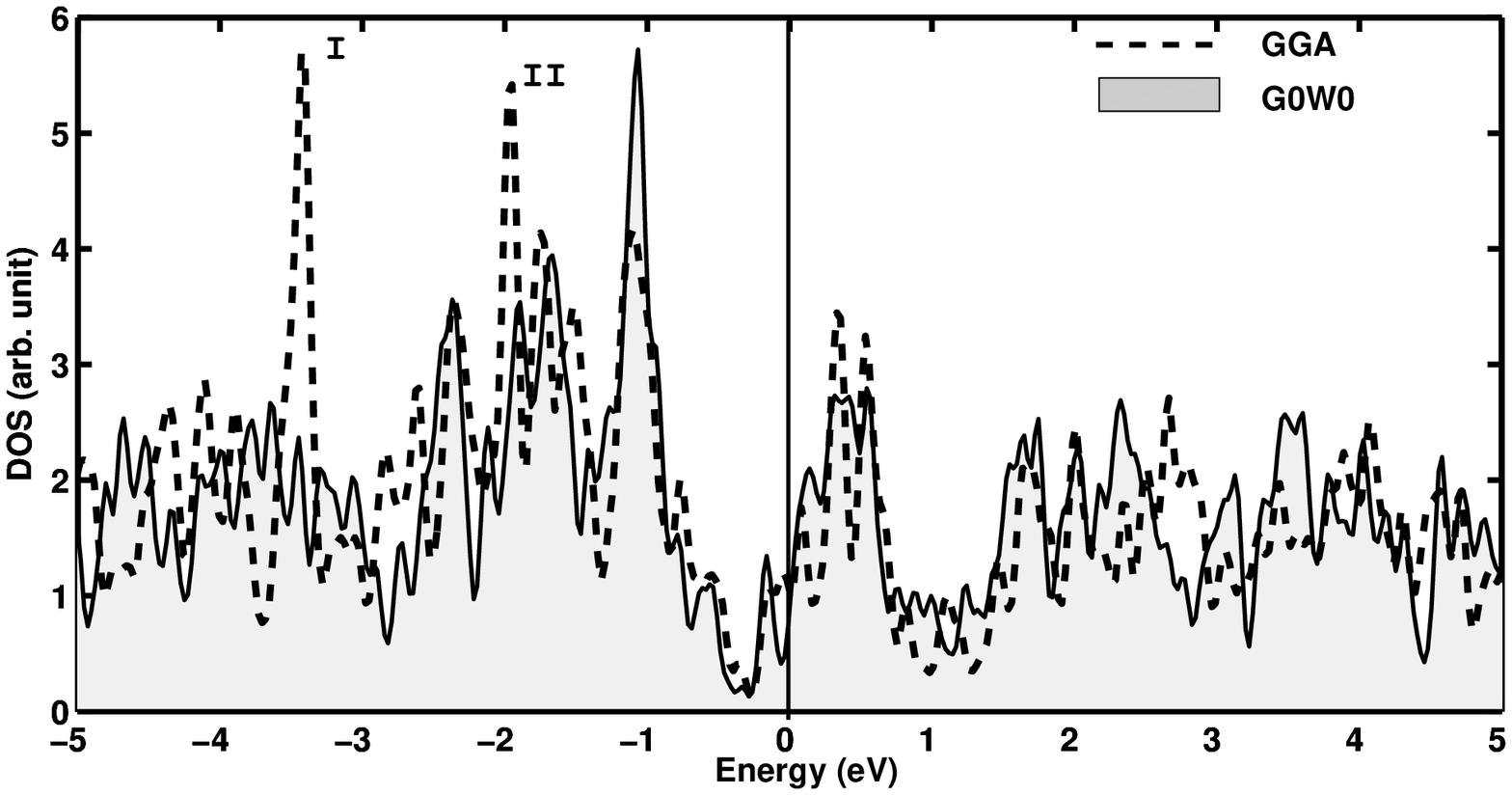}

\protect\caption{The left panel: The spin-polarised density of electron states for
$Fe_{3}Si$ in the interval of energies {[}-4eV, 2eV{]}. Since the
GGA and the GW approximations produce different Fermi energies, $\varepsilon_{F}^{GGA}(Fe_{3}Si)=7.88$
eV and $\varepsilon_{F}^{GW}(Fe_{3}Si)=8.44$ eV, the plots are aligned
for comparison by placing the zero in the energy axis of both plots
at Fermi energy.  The right panel: The DOS for $\alpha-FeSi_{2}$
in the energy interval {[}-5,5{]}eV with the same type of alignment
of the energy axes; $\varepsilon_{F}^{GGA}(\alpha-FeSi_{2})=9.34$
eV and $\varepsilon_{F}^{GW}(\alpha-FeSi_{2})=10.03$ eV. \label{fig:general DOSes} }
\end{figure}
\end{widetext}

\begin{widetext}

\begin{figure}[htb]
\includegraphics[width=0.4\linewidth,clip]{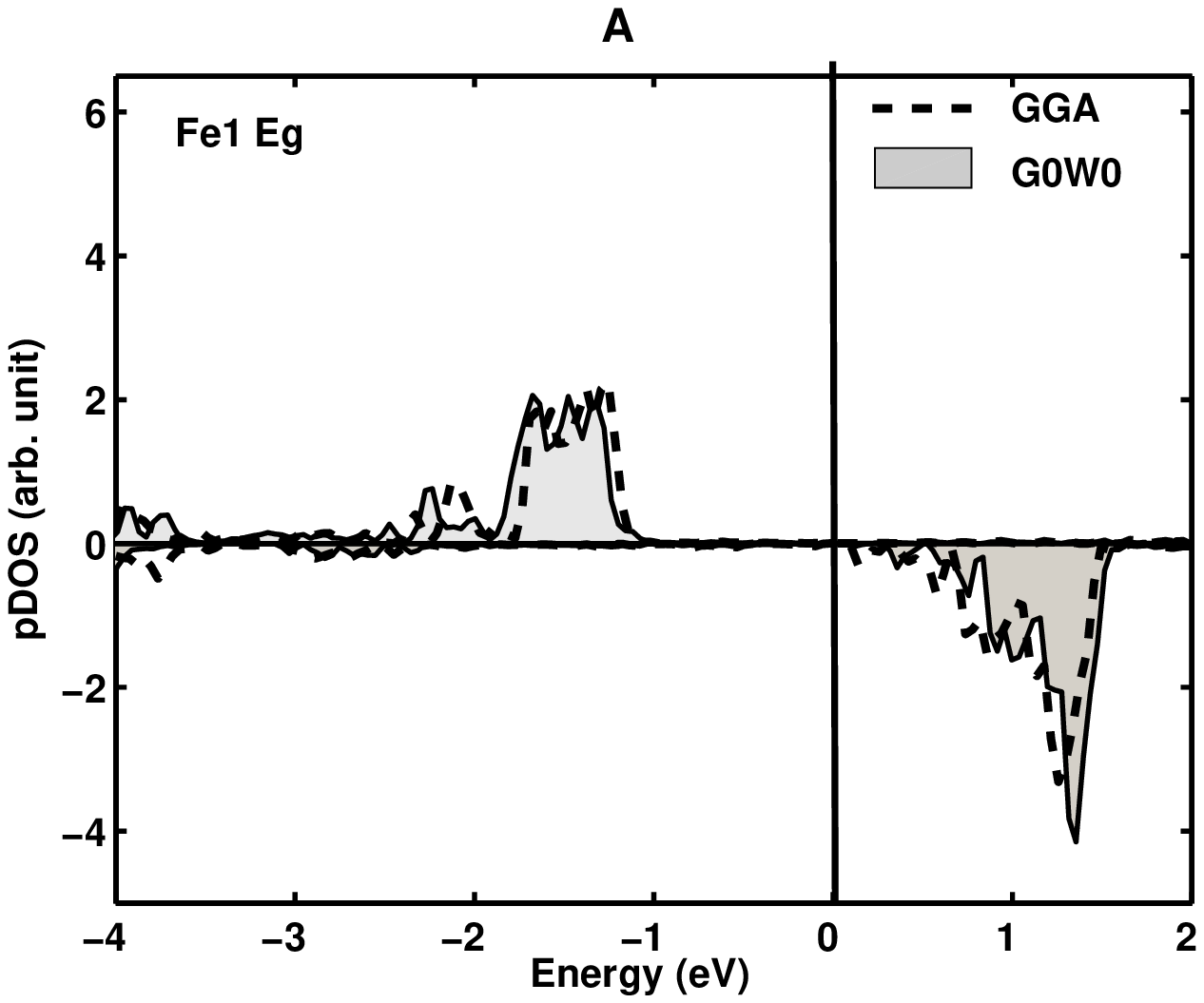} 
\includegraphics[width=0.4\linewidth,clip]{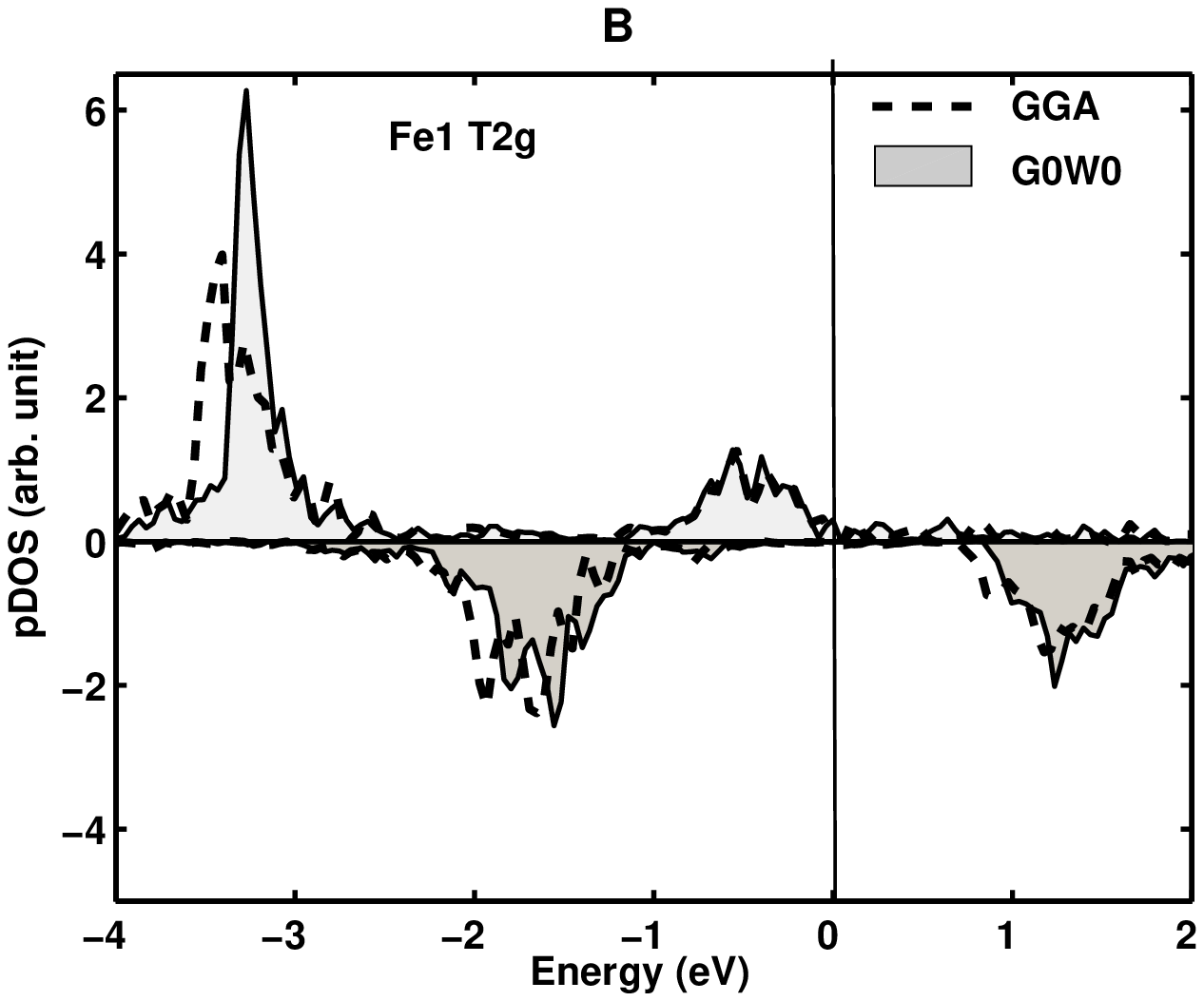} \\
\includegraphics[width=0.4\linewidth,clip]{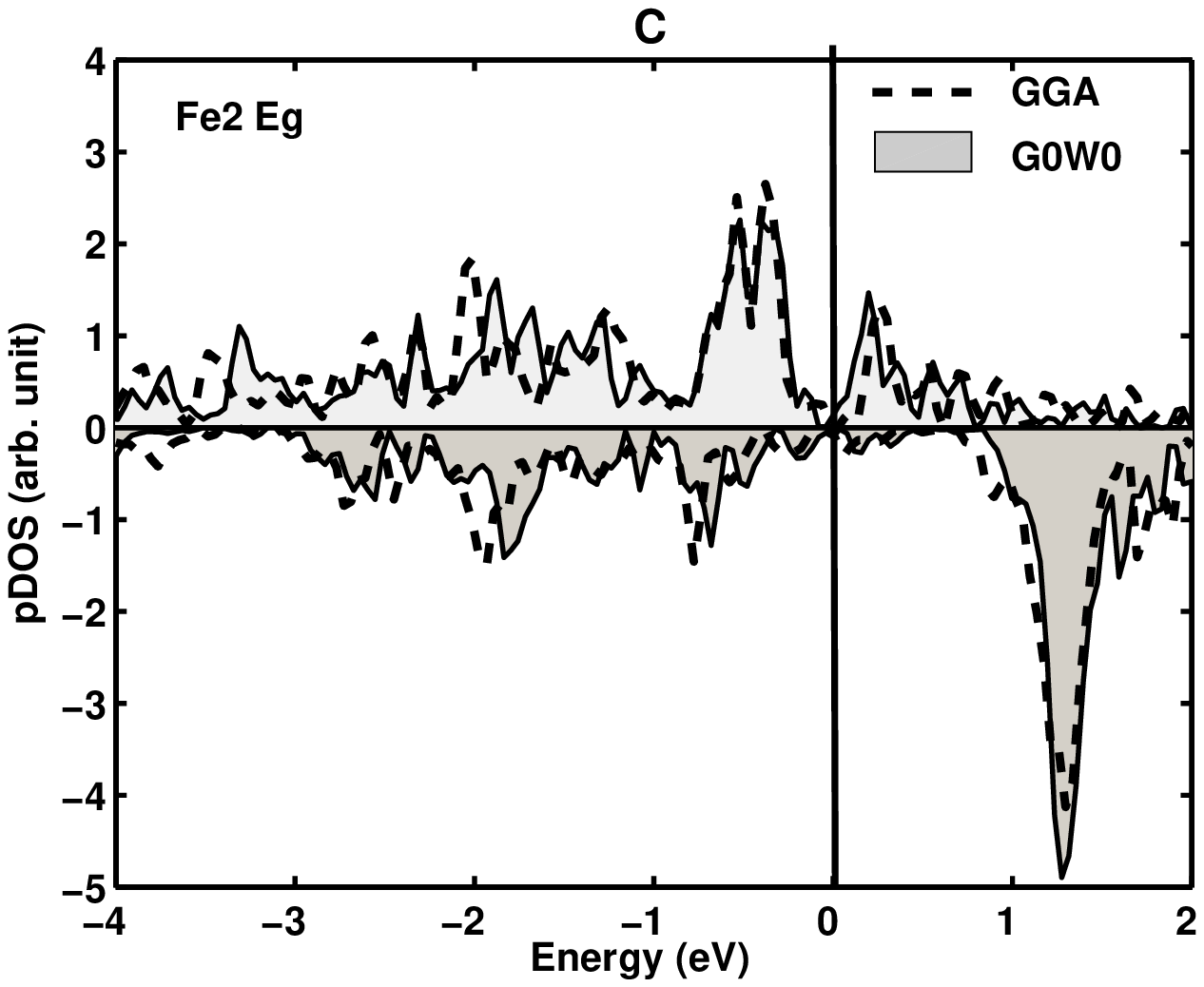}
\includegraphics[width=0.4\linewidth,clip]{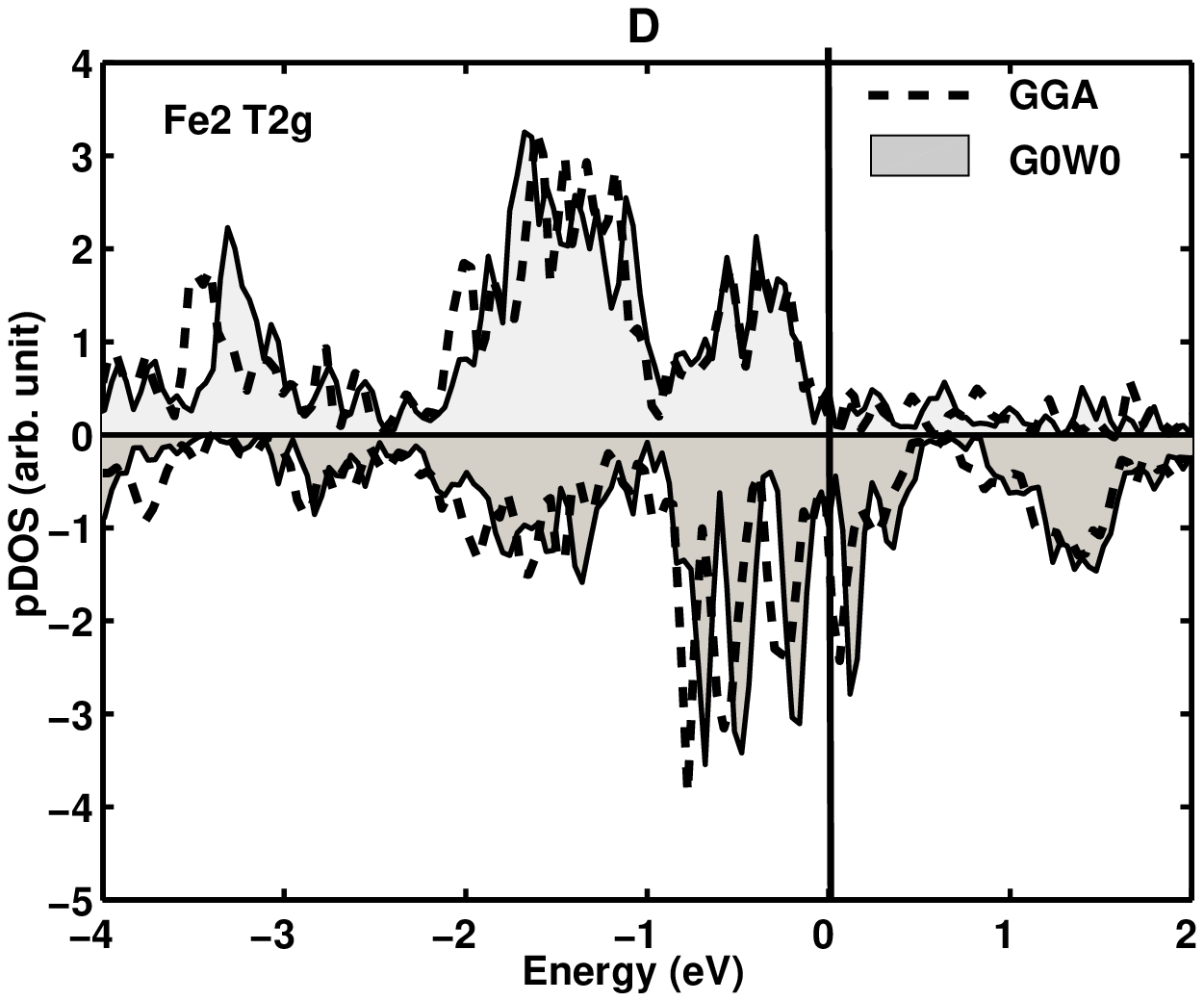}\\
\includegraphics[width=0.4\linewidth,clip]{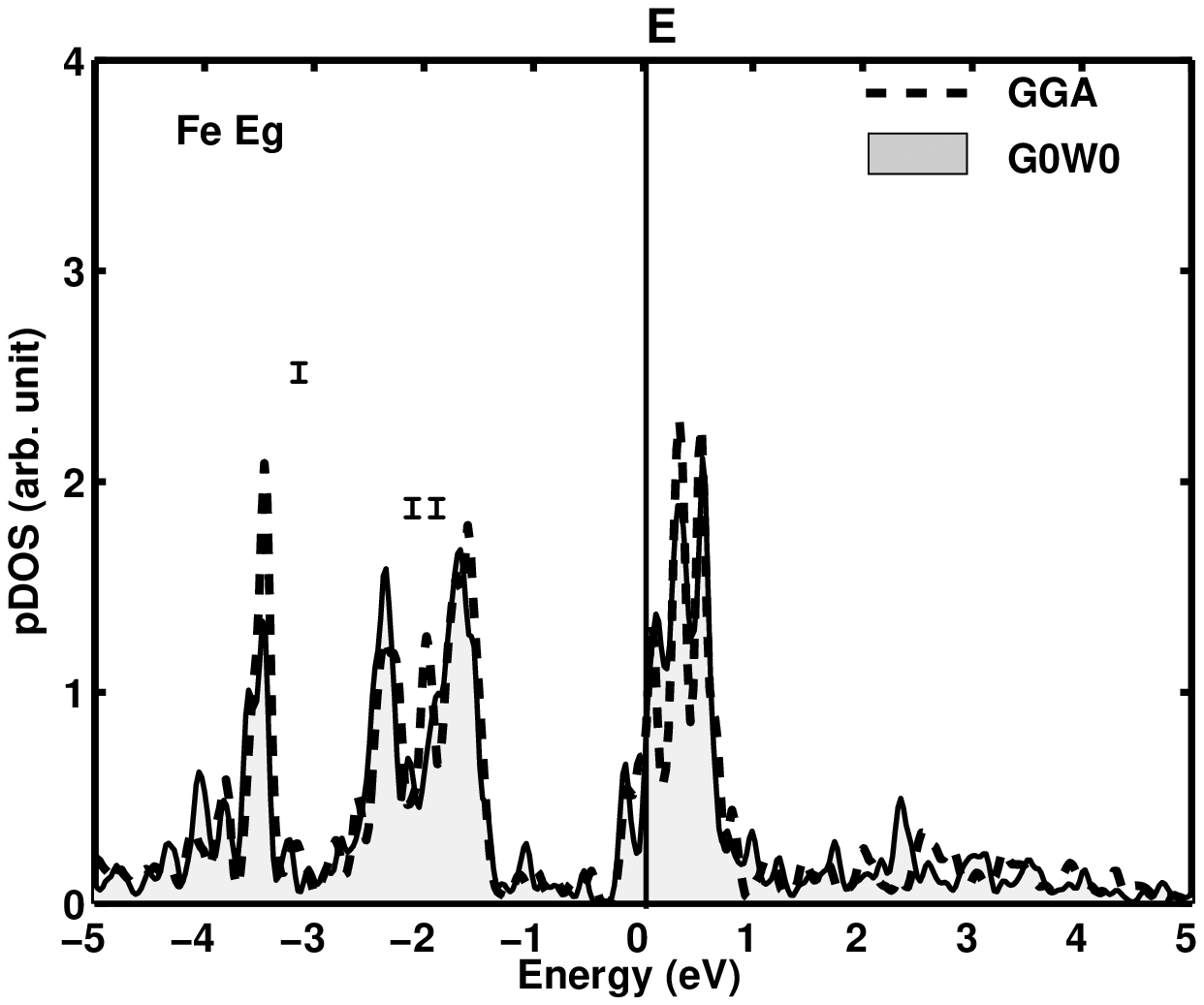}
\includegraphics[width=0.4\linewidth,clip]{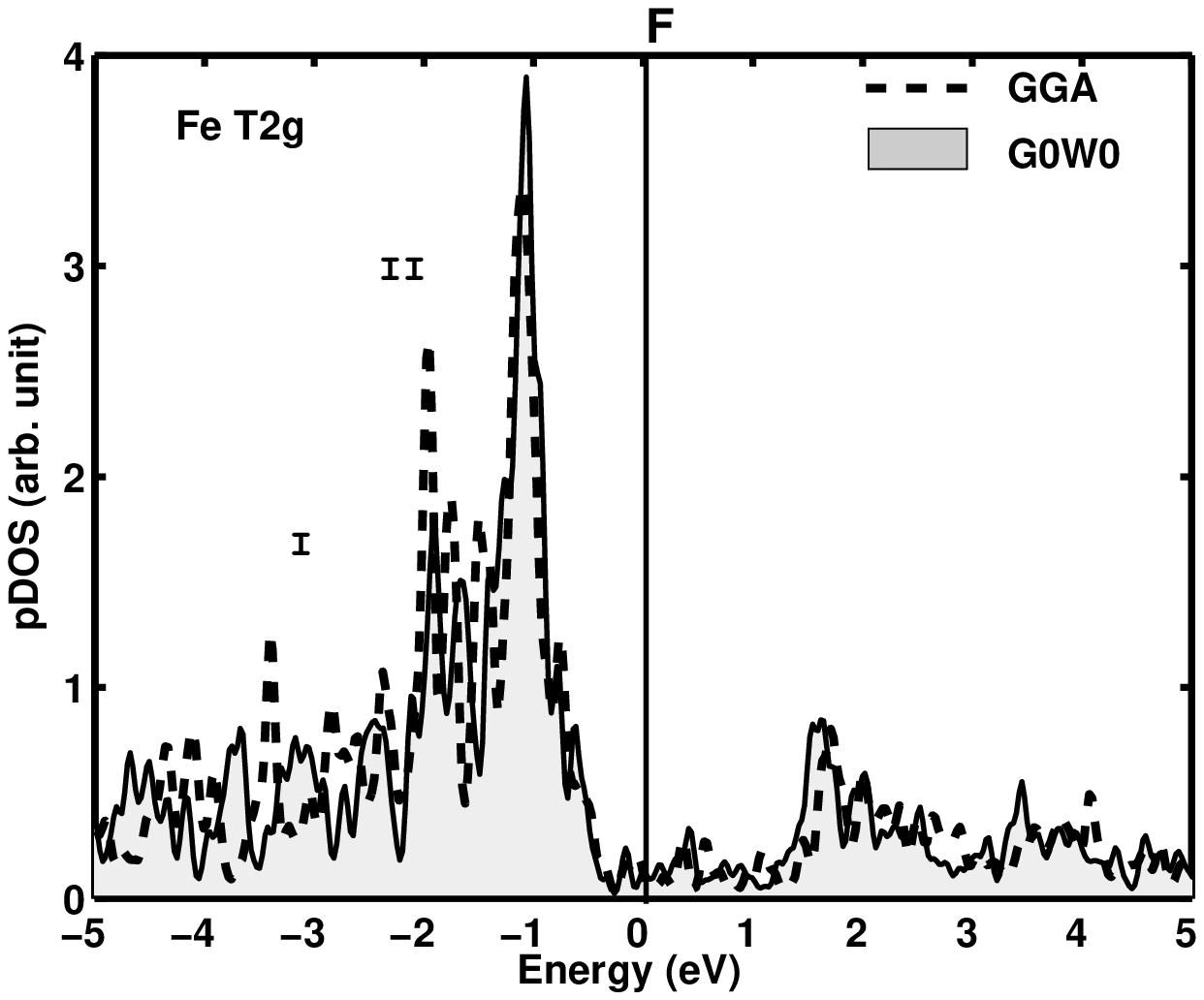}\\

\protect\caption{The partial spin-polarised $d$-electron DOS for $Fe_{3}Si$ (upper
four panels, (A),(B),(C),(D)) and $\alpha-FeSi_{2}$(lower two panels,
(E),(F)) \label{fig: partial-d-DOS}. Left panels display the contribution
to DOS from $d_{z^{2}}-$ and $d_{x^{2}-y^{2}}-$ states ($E_{g}$),
while the right ones show the contribution from $d_{xy}-$ , $d_{xz}-$
and $d_{yz}-$ states ($T_{2g}$). }
\end{figure}

\end{widetext}

Fig.(\ref{fig:general DOSes}) displays comparison of the GGA and
GW densities of electron states (DOS) for $Fe_{3}Si$ and $\alpha-FeSi_{2}$.
The GGA part of results coincides with previous calculations of $Fe_{3}Si$
\cite{key-24}, \cite{key-26}, \cite{key-27} and\textbf{ }$\alpha-FeSi_{2}$\cite{key-23}\textbf{,}\cite{key-26}\textbf{,}\cite{key-28}.
The general features of the DOS in both compounds and approximations
are that the bands in the interval$(-5,+5)$ eV around Fermi energy
are formed by the $d$-electrons of iron with a slight admixture of
$s-$ and $p-$electrons of $Si$ and $Fe$. The $Si$ valent
$s$- and $p-$electrons are delocalized in the wide energy
region with smeared maximum around $-4$ eV in both compounds. GWA
changes the intensities of the peaks mainly in the energy region deeply
under Fermi surface, but the changes in $Fe_{3}Si$ and $\alpha-FeSi_{2}$
are different. If  in the
GGA DOS of the $Fe_{3}Si$  the peak located at $E\sim-3.5$ eV is shifted by GWA for about $0.5$ eV and
made sharper, the GGA peaks in $\alpha-FeSi_{2}$ DOS in approximately
the same energy region (I, II in  bottom panel of Fig.(\ref{fig:general DOSes}))
is washed out within the GW calculations. There is a difference in
the GWA vs GGA changes in DOS for spin ``up''(majority) and ``down'':
while the GW spin-majority states remain almost untouched compared to GGA, the spin-minority
GWA peaks are shifted towards the Fermi energy. 
As was mentioned above, different chemical surroundings of the $Fe$
atom positions, the cubic one for $Fe^{(1)}$ by $Fe^{(2)}$ atoms
and the tetrahedral one for the $Fe^{(2)}$ atoms by the $Fe^{(1)}$
and $Si$ atoms as nearest neighbors reflect themselves in different
behavior of partial $d$-electron DOS. It is illustrated in Fig.(\ref{fig: partial-d-DOS}).

The general shape of partial $d$-DOS is not changed by GWA. The
$d$-states of $E_{g}$ symmetry of $Fe^{(1)}$ as well as $Fe^{(2)}$
do not contribute into DOS at the Fermi level in both in LDA and GWA
(panel (A), (C), Fig.(\ref{fig: partial-d-DOS})). As seen from Fig.(\ref{fig: partial-d-DOS}),
panels (A),(B), the contribution to the magnetic moment on $Fe^{(1)}$
from $E_{g}$ -orbitals (positive DOS in (A)) compensated by the contribution
from $T_{2g}$-orbitals ( the negative middle peak in (B)). This means
that the $T_{2g}$-orbitals are responsible for formation of the large
quasi-localized magnetic moment at $Fe^{(1)}$ atoms. It is also interesting
that the usual splitting of the $d$-shell into $T_{2g}$-
and $E_{g}$- symmetries is violated here by the contribution from exchange interaction:
 as seen from (A) and (B) panels the $E_{g}$-peak is in-between two $T_{2g}$-peaks. A different
behavior is demonstrated by $d$-electrons of $Fe^{(2)}$: its $d$-DOS
is spead in a wide region of energies. The $d$-electrons of both
$T_{2g}$- and $E_{g}$- symmetries contribute to formation of magnetic
moment of $Fe^{(2)}$. The delocalization of $Fe^{(2)}$ $d$-electrons
reflects itself in the smaller moment than the one on $Fe^{(1)}$.
The $d$-DOS of iron in $\alpha-FeSi_{2}$, where $Fe$ atoms also
have $Si$ atoms as neighbors, displays behavior similar to $Fe^{(2)}$
partial DOS for $d$-electrons in $Fe_{3}Si$ . However only these
delocalized electrons are not able to form magnetism in $\alpha-FeSi_{2}$.
Indeed, the criterium for magnetism formation for the delocalized
electrons is much harder to fulfil than for the case of localized
electrons whose role is played by $T_{2g}$- electrons of $Fe^{(1)}$
in $Fe_{3}Si$ . The absence of magnetism in $\alpha-FeSi_{2}$ is easy to understand
 on the basis of well-known Stoner's model 
for a magnetism of the delocalized electrons:  
 the criterium $Jg(\varepsilon_F)>1$ is not fulfilled since the density
of electron states $g(\varepsilon_F)$  at the Fermi energy $\varepsilon_F$
 is too small (here  $J$ is exchange integral
between delocalized electrons). An alternative mechanism of the magnetism supression  would be a formation 
of the low-spin  state within the localized d-electron picture. This state could be formed 
if the crystal-field splitting of the $d$-shell 
was stronger than the Hund-exchange one. However, the density of $d$-electron states does 
not contain bright peaks which might be interpreted as former $d$-levels splitted in the crystal field.
   Thus, one can conclude that if GGA and GWA are good approximations for $\alpha-FeSi_{2}$,
 the key mechanism of the magnetism destruction
in this compound  is the delocalization of $d$-electrons.  

The most pronounced changes in GWA compared to GGA  are experienced by $T_{2g}$ electrons. 
It is illustrated on Fig.(\ref{fig: partial-d-DOS}), panels (E,F) for $\alpha-FeSi_{2}$: two
peaks (I and II) seen in the GGA DOS which have been mentioned above
and washed out in the GWA are formed namely by $T_{2g}$ electrons.
The same is valid for the ``down'' spin \textbf{$T_{2g}$} states
in the vicinity of the Fermi level in $Fe_{3}Si$ (panel (D) in Fig.(\ref{fig: partial-d-DOS})).
At the same time the well-expressed localized peaks formed by $E_{g}$ orbitals
remain intact. This is understandable: the GWA is the approximation
which takes into account the effects of screening which is provided
namely by the delocaized electrons. These effects are taken into account
within the GGA and GWA differently and, therefore,
one can expect that the difference will be more pronounced for these
types of electrons.

\subsection{Comparison of GGA and GW band structures and analysis of the spectral
weights\label{sub:Compar-of-GGA_GW}}

\begin{figure}[htp]
\includegraphics[width=0.9\linewidth,clip]{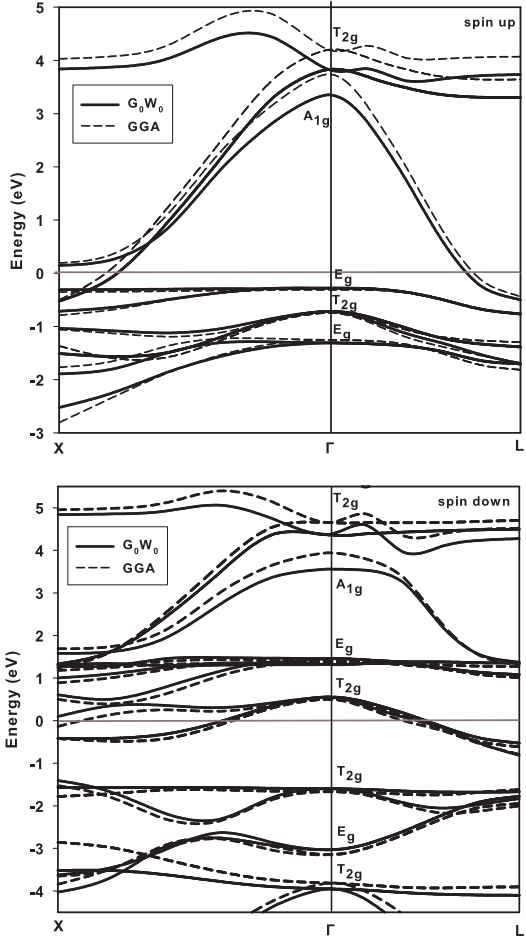} 

\protect\caption{\label{fig:Fe3Si_iBands}The GGA and GW spin-up (top) and spin-down (bottom) bands for $Fe_{3}Si$.
Zero in the energy axis of GGA and GWA plots is chosen at corresponding Fermi energies.  }
\end{figure}

\begin{figure}[floatfix]
 
%\begin{figure}[htp]
\includegraphics[width=1.0\linewidth,clip]{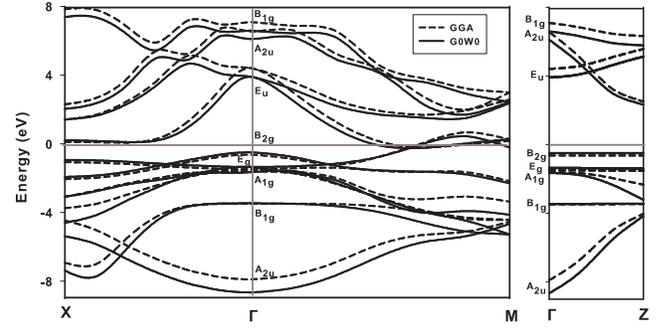}

\protect\caption{\label{fig:alphaFeSi2_bands}The band structure of $\alpha-FeSi_{2}$;
zero in the energy axis of GGA and GWA plots is chosen at corresponding
Fermi energies.  }
\end{figure}

The Kohn-Sham band structure calculated within GGA does not differ
from the known results for $Fe_{3}Si$ \cite{key-24}, \cite{key-27}
and for $\alpha-FeSi_{2}$ \cite{key-23},\cite{key-28},\cite{key-29}.
Here we compare the GGA and GW bands in some of symmetric directions.
Figs.(\ref{fig:Fe3Si_iBands}) and (\ref{fig:alphaFeSi2_bands}) show the
band structure for $Fe_{3}Si$ in the directions $\Gamma X$ and $\Gamma L,$
and for $\alpha-FeSi_{2}$ in the directions $\Gamma X$ , $\Gamma M$
and $\Gamma Z$, where $\Gamma=(0,0,0)$,   $X=\left(2\pi/a\right)(1,0,0)$,  $L=\left(\pi/a\right)(1,1,1)$,
$M=\left(2\pi/a\right)(1,1,0)$,  $Z=\left(2\pi/c\right)(0,0,1).$ 

%\begin{widetext}

%\begin{figure}[htp]
%\includegraphics[width=0.9\linewidth,clip]{Fig4b.eps}

%\protect\caption{\label{fig:Fe3Si_iBands-1}The GGA and GW spin-down bands for $Fe_{3}Si$.
%Zero in the energy axis of GGA and GWA plots is chosen at corresponding
%Fermi energies.  }
%\end{figure}
%\end{widetext}

We report here the results of the comparison only for the part of
the GGA and GWA band structures which are within several electron-volts
vicinity of the Fermi energies (remind that GGA and GWA generate different
Fermi energies, see the capture to Fig.(\ref{fig:general DOSes})).
The bands are named in accordance with their symmetries in the  $\Gamma-$point.
The closest to the Fermi energy three filled spin-up bands for the $Fe_{3}Si$
in Fig.(\ref{fig:Fe3Si_iBands}, left) near $\Gamma-$point, 
the dublet $E_{g}$ and the triplet $T_{2g}$,
are formed by the $d$-electrons of $Fe$ atoms. First empty band
($A_{1g}$) near $\Gamma-$point is formed by the $s$-states of both
$Fe$ and $Si$ atoms.

The GGA and GWA band structures for $\alpha-FeSi_{2}$ are shown in
Fig.(\ref{fig:alphaFeSi2_bands}). Here the closest to the Fermi energy
filled bands are formed by the $d$-orbitals of $Fe$ atoms near $\Gamma-$point
($D_{4h}$ group) are $B_{1g}$( $d_{x^{2}-y^{2}}$), $A_{1g}$ $(d_{z^{2}})$,
the dublet $E_{g}$ ($d_{xz}$,$d_{yz}$) and $B_{2g}$( $d_{xy}$). The
lowest shown band ($A_{2u}$) is formed by the $s$-electrons of $Fe$
and $p$-electrons of $Si$. Again we observe the same tendency: namely
the delocalized states, in this case, $s$- and $p$-states, show
the largest difference in GGA and GWA. If the band formed by $d$-electrons
close to the Fermi energy remain almost untouched, the\textbf{ }lowest
$sp$ band is shifted in GWA by $\sim1$ eV. First empty band ($E_{u}$)
near $\Gamma-$point is formed by the $p$-states of both $Fe$ and
$Si$ atoms. In general the GGA {\it vs} GWA shift is around half of
electron-volt for the excited states, while the band shape remains
the same. As seen at the right panel of Fig.(\ref{fig:alphaFeSi2_bands})
in the $\Gamma Z$ direction, the purely $d$-bands are completely
flat, while the dispersion which arises near the boundaries is due
to the admixture of $s$ and $p$ states. Analogous admixture of $sp$
electrons is observed around the boundary points $X$ and $M$. This is
easy to understand since the motion along $\Gamma X$ direction in
the $k$-space corresponds to motion from the $Fe$ atoms plane to
the plane of $Si$ atoms in real space;  $Si$ atoms do not have $d$-electrons and
can admix the $s$- and $p$-states only.

Here we return to the question asked in Sec.(\ref{sub:BandStr_vs_SpectrW}),
which electrons mainly are responsible for the deviations of GWA from
GGA results. Let us consider the example of two $E_{g}$ and $T_{2g}$
 filled bands of quasiparticles for spin ``up'' shown in Fig.(\ref{fig:Fe3Si_iBands}). 
%\begin{widetext}

\begin{figure}[htp]
%\begin{tabular}{ l r }
\includegraphics[width=0.95\linewidth,clip]{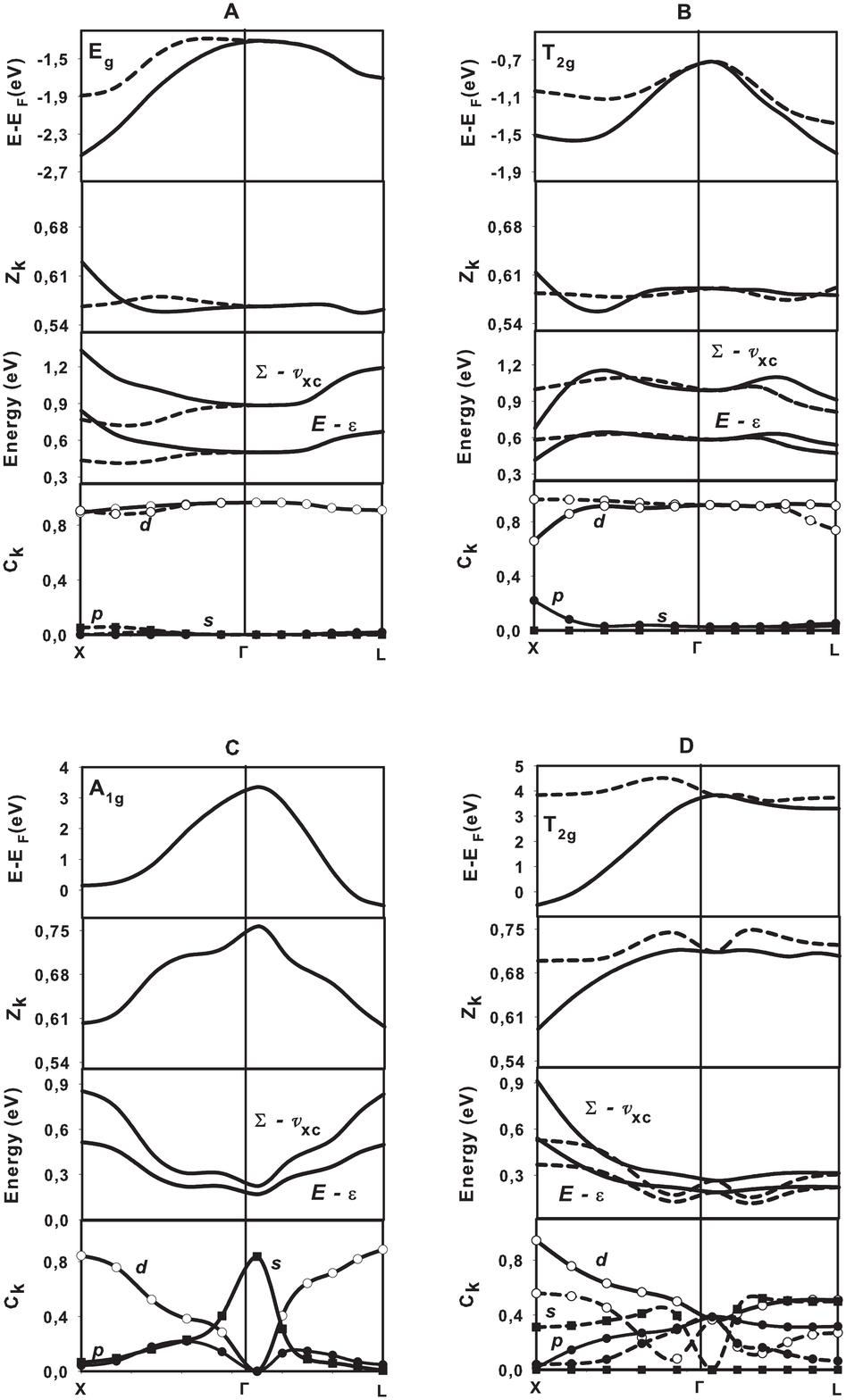} 
%
%\includegraphics[width=0.45\linewidth,clip]{Fig5a.eps}  \includegraphics[width=0.45\linewidth,clip]{Fig5b.eps} \\
%\includegraphics[width=0.45\linewidth,clip]{Fig5c.eps}  \includegraphics[width=0.45\linewidth,clip]{Fig5d.eps} \\
%\end{tabular}

\protect\caption{\label{fig: filledB:E_k,Z_k,C_Fe3Si} $Fe_{3}Si$ quasiparticle filled
bands (top panels), their spectral weights $Z_{k}$ (second from the
top), the $k$-dispersion of numerator and denominator of Eq.(\ref{eq:Z_kn})
(third from the top) and the character coefficients $C_{k}$(bottom
panels, see text). The letters label different characters: $d$ and
$p$ stand for the empty and filled circles correspondingly, the black
squares denote $s$ character. The filled bands are shown in (A) and
(B) panels, while (C),(D) display the empty ones. The dashed and solid
lines on (A) panel denote the non-degenerate bands $E_{g}$. The dashed
and solid lines on (B), (D) panel denote non-degenerate and double
degenerate bands $T_{2g}$. }
\end{figure}
%\end{widetext}

\begin{figure}[htp]
\includegraphics[width=0.95\linewidth,clip]{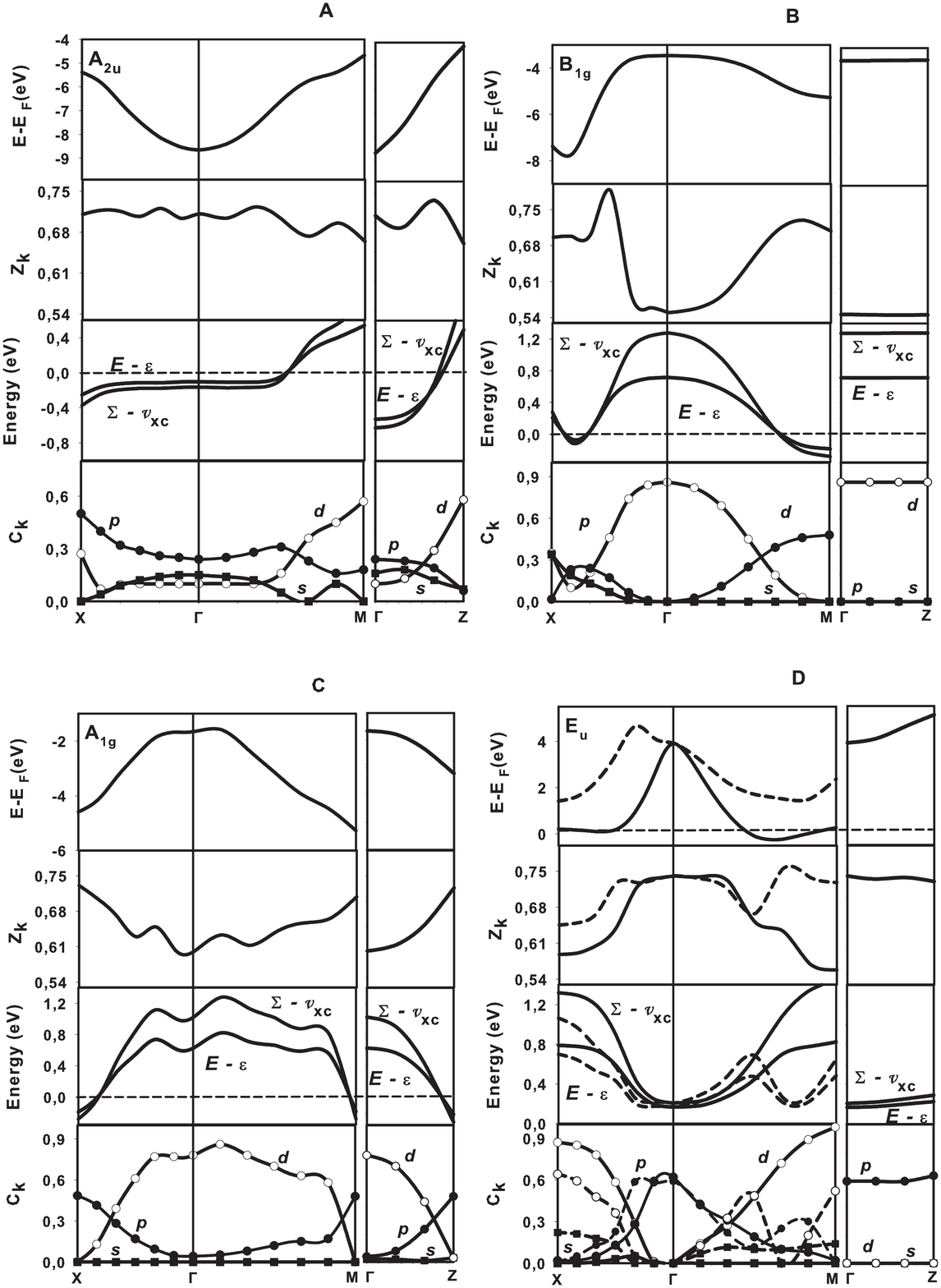}

\protect\caption{\label{fig:FeSi2_bands_Z_}The $\alpha-FeSi_{2}$ bands. The notations
are the same as in Fig.(\ref{fig: filledB:E_k,Z_k,C_Fe3Si}), but
the filled bands are imaged at the panels A,B,C, while the empty one
is shown at D panel.}
\end{figure}

First general feature seen from both Fig.(\ref{fig: filledB:E_k,Z_k,C_Fe3Si})
and Fig.(\ref{fig:FeSi2_bands_Z_}) is that the spectral weights $Z_{kn}$,
that within the Kohn-Sham approach are equal to one by construction,
within the GWA are strongly decreased. Second feature, somewhat surprising,
 is that the non-monotonous behaviour of $Z_{kn}$ arises everywhere
where an admixture of delocalized electron states is present. Indeed,
 one observes that near the $X$ point of the Brillouin
band it is the admixture of $s$- and $p$-electrons to $d$-states
for the band $T_{2g}^{(1)}$ of $Fe_{3}Si$ causes the changes in
the $Z_{kn}$; the difference  $Re\Sigma'-v_{xc}$ is changed faster
than $E_{kn}-\varepsilon_{kn}$ (see panel A at Fig.(\ref{fig: filledB:E_k,Z_k,C_Fe3Si})).
The picture for the empty bands is different: the $s$- and $p$-states
of $Si$ and $d$-states of $Fe$ are mixed in the center of the band,
whereas the contribution of the $d$-states is increased around boundary
points $X$ and $L$. The quasiparticle energies of the excited states
are lower than their Kohn-Sham counterparts. Again, the largest difference
is observed in the those parts of the energy spectrum where the contributions
from $s-$ and $p-$ states becomes significant.

%\begin{widetext}

%\end{widetext}
As seen from Fig.(\ref{fig:FeSi2_bands_Z_}), the spectral weights
$Z_{kn}$of the quasiparticles in $\alpha-FeSi_{2}$ also show a strongly
non-monotonous dependence on $k$. For example, the spectral weight
$Z_{k,B_{1g}}$ of the $B_{1g}$ band in the direction $\Gamma X$
(second from top panel in Fig.(\ref{fig:FeSi2_bands_Z_},B) shows
sharp changes in the interval $0.55<Z_{k,B_{1g}}<0.8$ due to much
faster dependence on $k$ of the difference $\left(\Sigma'-v_{xc}\right)$
 than of the $\left(E_{QP}-\varepsilon_{GGA}\right)$ one.
The lower part of the panel B in Fig.(\ref{fig:FeSi2_bands_Z_}) explains
the reason: again, the closer to the $X$ point the higher contribution
from the delocalized $s$-electrons. This confirms, thus, the general
tendency noticed above.

\section{Discussion and Conclusions}

The comparison of the band structures obtained in the {\it ab initio}
calculations within the VASP for $Fe_{3}Si$ and $\alpha-FeSi_{2}$ 
in $GGA$ and $G_0W_0A$ shows that in general
the bands shape is similar. The difference between GGA and GWA bands
becomes more pronounced in those parts of the Brillouin zone where
 the delocalized states give noticible contribution into quasi-particle energies.
 This observation is somewhat unexpectable since both approximations are designed for
description of well delocalized (Fermi-liquid like) electrons. 
There are at least two sources which can contribute to 
this difference. First is the fact that the standard GGA
is not free from the self-interaction while GWA takes into account
Fermi statistics by construction. Second source is that GGA and GWA are quite 
different approximations, as was discussed in Introduction.\\
The other astonishing moment following from GWA calculations is that
in spite of the fact that the electrons in both systems are well delocalized
the spectral weights $Z_{kn}$ of the quasiparticle bands $E_{kn}$
are decreased almost by a half.  One could 
assume that the reason for that is that the basis set is not sufficiently
large, however, the results are not changed with further increase of number of bands
which are taken into account. This means that the GWA indeed shifts the
remaining part of the weight to the incoherent part of excitations.

Both GGA and GWA band structures and, correspondingly, the density
of electron states, show that the $d$-electrons of those $Fe$ atoms
which have $Si$ nearest neighbors, namely, $Fe^{(2)}$ atoms for
$Fe_{3}Si$ and all $Fe$ atoms in $\alpha-FeSi_{2}$, are more delocalized
than the $d$-electrons of $Fe^{(1)}$ atoms in $Fe_{3}Si$ which
have only the other $Fe$ atoms as neighbors. The partial density
of states of $Fe^{(1)}$ $d$-electrons with $E_{g}$ and $T_{2g}$
symmetry in the $\Gamma$ point has well-expressed peaks, the positions
of which could be ascribed to a splitting in the crystal field. However,
this splitting does not correspond to the standard picture of the
quasiatomic levels
 $\varepsilon_{T_{2g},\sigma}^{0}d_{t\sigma}^{\dagger}d_{t\sigma}+\varepsilon_{E_{g},\sigma}^{0}d_{e\sigma}^{\dagger}d_{e\sigma},$
from which the bands are formed, the interactions renormalize these
``levels'' $\varepsilon_{T_{2g},\sigma}^{0}\rightarrow\varepsilon_{T_{2g},\sigma}$
in such a way that their sequence becomes
 $\varepsilon_{T_{2g},\uparrow}<\varepsilon_{T_{2g},\downarrow},\varepsilon_{E_{g},\uparrow}<\varepsilon_{T_{2g},\uparrow.}$\\
We performed the calculations within the one-shot GW approximation ($G_0 W_0$ ), and it is worth to notice why this approximation
 is preferable compared to the fully self-consistent GWA.
First, it often gives the results which  describe PES experiments better (see, {\it e.g.,} refs.\cite{key-32},\cite{key-33},\cite{key-34}). 
Second argument is connected with the nature of the self-consistency. Each term in the self-consistent perturbation theory (scPT) corresponds 
to whole series in the non-scPT. If we consider these corrections, we find that a part of corrections in each order
 of the non-scPT changes the matrix elements of the Coulomb interaction, actually, via corrections to the wave functions. The self-consistency loop changes the eigenfunctions 
which diagonalize the GW equation for the Green's functions, and, correspondingly, dresses the matrix elements of 
Coulomb interaction by the RPA (random phase approximation) graphs only.   All other contributions to the wave functions, which arise in the same orders  of non-scPT, 
and, correspondingly, to the matrix elements of Coulomb interaction, are not taken into account.  
However, in the case  of non-homogeneous electron gas in real-material, where the gellium-like  parts can coexist with strongly correlated 
liquid, the situation it not alway obvious.  The question how and to which terms the vertex corrections should be applied remains highly non-trivial. This have been demonstrated,
{\it e.g.}, in ref.\cite{key-35 },   where it was shown that the vertex corrections into effective interaction $W$ may improve the results while turning on the corrections 
in the self-energy may lead to unphysical quaiparticle dispersion. 
At last, the calculations within GWA  with the vertex corrections demand for much stronger computer resources than even scGWA. 
The ARPES experiments on iron silicides would be of great help in further understanding of these compounds and, possibly, could motivate 
more advanced theoretical approach to the problem.
\\

\section*{Acknowledgement}

The authors thank for support from RFBR No. 14002-00186, the President
of Russiun Federation Grants (NSH-2886.2014.2 and NSH-924.2014.2) and Physics Department of RAS program
"Electron correlations in systems with strong interaction".
The calculations were performed with the computer resources of NRC
``Kurchatov Institute'' (ui2.computing.kiae.ru). I.S. thanks A.Ruban for 
useful discussion.

\end{document}